\documentclass[
twocolumn,
amsmath,amssymb,
aps,
prb,
showpacs,
]{revtex4-1}
\pdfoutput=1

\usepackage{graphicx}

\begin{document}
\title{Quantum Coulomb gap in low dimensions}
\author{M. \surname{Pino}}
\affiliation{Departamento de F\'{\i}sica - CIOyN, Universidad de Murcia, Murcia 30071, Spain} 
\author{A.~M. Somoza}
\affiliation{Departamento de F\'{\i}sica - CIOyN, Universidad de Murcia, Murcia 30071, Spain} 
\author{M. Ortu\~no}
\affiliation{Departamento de F\'{\i}sica - CIOyN, Universidad de Murcia, Murcia 30071, Spain} 

\begin{abstract}
We study the single-particle density of states of one-dimensional and two-dimensional quantum disordered systems with long-range interactions.
We consider a $1/\sqrt{r}$ interaction in one dimension and a Coulomb interaction in two dimensions,
which produce linear gaps in the density of states in both cases. 
We focus on the strong localization regime where the localization length is small but non-zero. 
We use an exact diagonalization technique for small system sizes and a perturbation approach for larger sizes.
We find that, with both methods, the inclusion of a finite hopping contribution does not change the linear character of the gap, but reduces its slope, widening the gap.
\end{abstract}
\maketitle
\section{Introduction}\label{section:1}

Insulators can exhibit interesting transport properties. At very low temperatures
localized electrons jump via phonon-assisted hopping and the transition rates depend
exponentially on distance and energy. The typical hopping distance and energy are functions of the temperature and the 
mechanism is known as variable-range hopping.
The non-interacting version of this phenomenon was explained forty years ago by Mott [\onlinecite{Mott}].
When interactions between electrons are taking into account the problem becomes much more complicated.
In the case of Coulomb interactions, Efros and Shklovskii (ES) proposed an extension 
of the non-interacting case in which the constant density of states (DOS) of the non-interacting case is replaced by the DOS of the interacting one [\onlinecite{Efros_Shklovskii}]. 
This DOS presents a
gap, known as the Coulomb gap, around the Fermi level [\onlinecite{Pollak}]
which limits the hopping of electrons at low energies.
The shape of the Coulomb gap depends on the dimension of the system $d$ and is given by:
\begin{align}
\rho\left( \epsilon\right)\sim|\epsilon| ^{d-1},
\end{align}
for $d>1$. In this formula the energy $\epsilon$ is measured with respect to the Fermi energy. For $d=1$ there exist logarithmic corrections to the DOS [\onlinecite{Efros1d}].
One implicit approximation performed in the ES argument is 
that weak quantum effects do not alter the shape of the DOS [\onlinecite{Efros76}]. 

In this work we study the effects of a small transfer energy contribution $t$ in the shape of the Coulomb gap. 
To generalize the definition of the classical Coulomb gap to quantum systems we employ the local DOS.
The particle contribution $\rho_{i}^{p}(\epsilon )$, and the hole contribution $\rho_{i}^{h}(\epsilon)$ to this local DOS at site $i$ are [\onlinecite{Economu}]:
\begin{align}
 \rho^{p}_{i}(\epsilon )=\sum_{\alpha}\delta\left(\epsilon -E^{p}_{\alpha}+E_{0}\right) \left|\langle \psi^{p}_{\alpha} |c_{i}^{\dag}
|\psi_{0}\rangle\right| ^{2},\label{eq:dp}\\
 \rho^{h}_{i}(\epsilon)=\sum_{\beta}\delta\left(\epsilon +E^{h}_{\beta}-E_{0}\right) \left|\langle \psi^{h}_{\beta} |c_{i}|\psi_{0}
\rangle\right| ^{2}, \label{eq:dh}
\end{align}
where $E_{0}$, $|\psi_{0}\rangle$ are the eigenenergy and eigenstate of the ground state in the grand canonical ensemble
which we assume contains $n$ electrons. The energy of an eigenstate $\alpha$ with an electron added to the ground states is denoted by $E^{p}_{\alpha}$
and its corresponding wave function by $|\psi^{p}_{\alpha}\rangle$. For the energy and the wave function of an eigenstate $\beta$ with $n-1$ electrons we use $E^{h}_{\beta}$
and $|\psi^{h}_{\beta}\rangle$ respectively.
We notice that $\rho^{p}_{i}(\epsilon )$ and $\rho_{i}^{h}(\epsilon )$ have been defined in such a way that a positive energy is needed to add a particle 
while a negative one is necessary to subtract a particle. 
Using these particle and hole contributions, we express the average of the local DOS:
\begin{equation}
\rho(\epsilon)  = \left\langle\rho_{i}^{p}(\epsilon)+ \rho_{i}^{h}(\epsilon)\right\rangle, \label{eq:dt}
\end{equation}
where the brackets mean average over different sites $i$ or disorders.
This expression reduces to the classical DOS when quantum effects are ignored.
In the following, we refer to $\rho(\epsilon)$ just as the DOS. 

An intuitive argument to describe the changes in the Coulomb gap due to quantum effects is given in Ref.\ [\onlinecite{Epperlein97}].
In the classical case, the restriction on the DOS near the Fermi level is due to long-range interactions.
If quantum effects only produce a small increase of the localization length, the nature of those long-range interactions
should remain the same at scales larger than the localization length and so at energies (in units of $e^2$) smaller than its inverse. 
Then, one can conclude that the shape of the DOS close enough to the Fermi energy should not change. 
This picture agrees with the experimental results on tunneling conductance [\onlinecite{Massey_Lee_Shklovskyi},\onlinecite{experimental_gap2}], 
although the relation between the DOS and the tunneling conductance is not yet fully understood [\onlinecite{CuOr92}].
We want to emphasize that this argument does not seem to depend on the dimensionality of the system and the same behavior may be expected for any dimension.

The first attempt made to include quantum effects on the Coulomb gap was performed by Vignale and coworkers [\onlinecite{Vignale86}]. 
They compute the DOS for non-interacting electrons. But, they chose an on-site energy distribution given by the classical Coulomb gap 
instead of being randomly distributed. Using the results obtained for this calculation, they claimed that the 
Coulomb gap is unstable under the inclusion of arbitrarily small quantum effects in two-dimensional (2D) systems. 
In three dimensions, they found that the gap becomes narrower and quantum effects produce a non-zero density at the Fermi level.
After that, Vignale [\onlinecite{Vignale87}] used a Hartree-Fock based approximation 
to obtain that the gap narrows for 2D and three-dimensional (3D) systems.
In Ref.\ [\onlinecite{Li93}] a coherent potential approximation was employed to conclude that the gap narrows 
but the DOS at the Fermi energy remains negligibly small when weak quantum effects are considered in two and three dimensions.
Schreiber et al. [\onlinecite{Epperlein97}] studied the 3D case within a Hartree-Fock based method. Their results show that 
weak quantum effects narrow the gap and  are compatible with a zero DOS at the Fermi energy. 
The narrowing of the gap with a finite $t$ seems to be the natural tendency at least in 3D systems, 
since the gap must be logarithmically small in the extended phase [\onlinecite{AtAr}].
However, Jeon et al. [\onlinecite{Lee99}] applied the same method to one-dimensional (1D) and 2D systems and their results show that  
the slope of the classical gap near the Fermi level decreases upon including weak quantum effects.
This is in contradiction with previous results [\onlinecite{Vignale86},\onlinecite{Vignale87},\onlinecite{Li93}] in two dimensions.
We note the  lack of agreement on whether weak quantum effects widen or narrow the Coulomb gap 
and on whether they produce or not a finite DOS at the Fermi level.
Neither is clear if  quantum corrections depend on the dimensionality of the system.

We study the DOS given by Eq.\ (\ref{eq:dt}) for systems composed by many electrons when disorder 
and long-range interactions are important but quantum effects are weak.
Our main goal is to study 2D systems with Coulomb interactions, but we also analyze 1D systems with  a $v(r)=1/\sqrt{r}$ interaction.
This interaction produces a linear gap in the classical DOS of 1D systems, the same as the Coulomb interaction in the 2D case.
This linear gap is easier to analyze numerically than the logarithmic gap produced by
a Coulomb interaction in one dimension.
Furthermore, we will see that our results for 1D systems show the same tendency that the ones found in Ref.\ [\onlinecite{Lee99}] for Coulomb interactions.

The structure of this paper is the following.
In section\ \ref{section:2}, we introduce the model and the basic numerical tools which shall be used.
After that, in section\  \ref{section:3},
we present results obtained with exact diagonalization.
In section\ \ref{section:4},  a perturbative analysis of the DOS is performed and the results of this analysis are used to compute the DOS 
for much larger system sizes than those obtained by exact diagonalization.
Finally, our results are briefly summarized and discussed in section\ \ref{section:5}.

\section{Model and method}\label{section:2}

We study a spin-less fermionic system in a regular lattice with long-range interactions in which translational symmetry is broken by an on-site
random potential and particles can tunnel to nearest neighbors sites. We use a tight-binding Hamiltonian:
\begin{align}\label{eq:hamiltonian}
  H     =&\sum_{i} \left(\phi_{i}-\mu\right) n_{i}+
V \sum_{i\neq j}\frac{\left({n}_{i}-K\right) \left({n}_{j}-K\right)}{r_{ij}^{\alpha}} \nonumber \\
         & -t\sum_{<i,j>} c_{i}^{\dag} c_{j},
\end{align}
where  $\phi_{i}$ is a random site energy with an uniform distribution in the interval $[-W/2,W/2]$, $t$ is the hopping parameter, $V$ is 
the strength of interaction, $\mu$ is the chemical potential and  $K$ is the compensation constant. The parameter $\alpha$ 
is chosen depending on the dimension as explained in the introduction.
That is, $\alpha=1/2$ is used in 1D and $\alpha=1$ for 2D systems.  
In our computations we implement periodic boundary conditions in all the directions 
and the interaction between two sites is set by the minimum-image convention.
We take an interaction energy $V=1$, which fixes the energy scale, and 
we consider a disorder energy $W=2$.
We are mainly interested in a regime where the localization length is small compared with the system size,
but non zero. We use a compensation constant $K$ and chemical potential $\mu$ depending 
on the filling to ensure the neutrality of the system.
Most of our work has been performed at half filling for which $K=1/2$ and $\mu=0$. 
For this value of $K$ the DOS is symmetric with respect to the Fermi energy. We shall take advantage of this property
to only show half of the DOS when presenting results.
We also compute in section\ \ref{section:4} the DOS at one third filling. In this case, we  set
$K=1/3$ and $\mu=-1/3$.

We analyze the two alternative approaches employed to obtain the DOS, Eqs.\ (\ref{eq:dp}) and (\ref{eq:dh}).
The first one is based on exact diagonalization. The largest system sizes considered with this method are $L=24$ for one dimension and 
$N=4\times 6$ for two. 
In order to compute the DOS via exact diagonalization, we proceed as follows.
We search for the ground state in the grand canonical ensemble which we assume contains $n$ particles. 
As we will discuss later, this ensemble minimizes finite size effects.
We are interested in an energy region close enough to the Fermi level, 
so we only calculate the $S$ smallest energy levels for the cases corresponding to $n+1$ and $n-1$ particles, with $S\sim 50$.
Such a partial diagonalization is carried out using ARPACK 
open library [\onlinecite{ARPACK}] which are based on a variant of the Lanczos algorithm and allows to efficiently obtain a few eigenvectors and its eigenenergies.
In Ref.\ [\onlinecite{Israel}], we provide a thorough study of the validity of the method depending on the number of eigenstates considered.
We run this procedure for at least 3000 realizations of disorder to obtain an average of Eqs.\ (\ref{eq:dp}) and (\ref{eq:dh}).
Finally, by working in the grand canonical ensemble and choosing the minimum energy eigenstate
for all the occupations, we ensure that the electron energies are always positive (hole energies negative) and this
avoids an artificial filling of the DOS at the Fermi energy.
An analogous problem in the classical limit is discussed in Ref.\ [\onlinecite{Efros_computation}].

In the second approach we use a numerical algorithm to obtain the classical ground state, i.e. the lowest energy configuration of the $t=0$ limit and 
a fairly complete set of low-energy configurations, needed to recalculate the new ground state in 
the presence of quantum effects as will be explained in section \ref{section:4}.
To compute these classical states, we repeatedly start from states chosen at random and relax each sample using a local search procedure [\onlinecite{eg-classic3}].
In an iterative process, we look for configurations of lower 
energies differing by one-electron or compact two-electron jumps
and always accept the first of such states found. The procedure 
stops when no lower energy neighboring states exist,
which insures stability with respect to all one-electron jumps
and compact two-electron jumps. We then consider the set of
metastable states found by the process just described and look
for the sites which present the same occupation in all of them.
These sites are assumed to be frozen, i.e. they are not allowed
to change occupation, and the relaxation algorithm is now applied
to the unfrozen sites. The whole procedure is repeated
until no new frozen sites are found with the set of metastable
states considered. The lowest energy configuration of the set
of metastable states is consider to be the classical ground state.
The set of low-energy states is completed by generating all the
configurations that differ by one or two electrons 
transitions from any of the metastable configuration state found before.

\section{Exact diagonalization}\label{section:3}

\begin{figure}[t!]
\begin{center}
\includegraphics[width=1.0\linewidth]{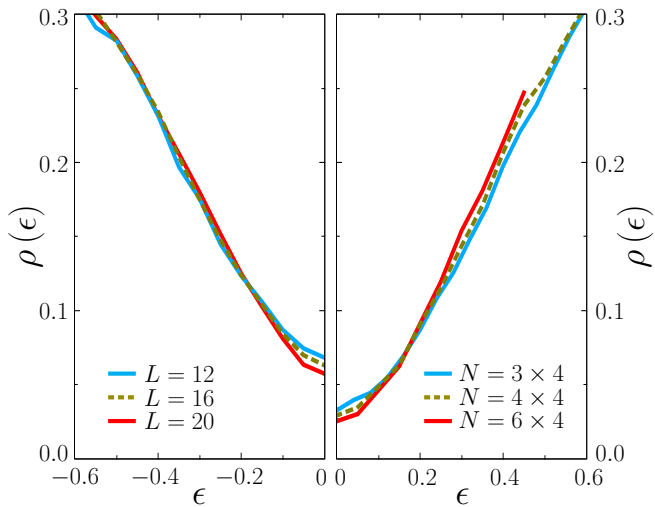}
\caption[]
 {(Color online). The DOS as a function of energy for 1D (left) and 2D (right) systems. 
The hopping is $t=0.2$ and the sizes are specified in each panel. 
}\label{fig:Fig1}
\end{center}
\end{figure}

As the sizes involved in the exact diagonalization are small, let us start with a finite size analysis to check that 
valid conclusions still can be drawn.
In Fig.\ \ref{fig:Fig1} we show the DOS for hopping parameter $t=0.2$ in one (left) and two dimensions (right). 
The sizes employed are $L=12,\ 16,\ 20$ in one dimension and $N=4\times 3,\ 4\times 4,\ 4 \times 6$ in two dimensions.  
The symmetry of the DOS with respect to the Fermi energy at 
half occupation is taken into account and only half of the DOS is represented.
In the classical case, finite size effects produces a filling of the DOS at the Fermi energy 
depending on the linear size of the system $L$ as $\rho\left(\epsilon=0\right)\sim 1/L$ for both 1D and 2D systems [\onlinecite{Efros_computation}].
As in this classical case, the results shown in Fig.\ \ref{fig:Fig1} indicates that finite size effects for small hopping are important for energies close to the Fermi level.
But they are not so important at intermediate energies inside the Coulomb gap.
Indeed, we can observe in both panels appearing in Fig.\ \ref{fig:Fig1} 
that there is a region around the middle of each branch of the gap where the changes in the DOS are small for
the sizes considered. We take advantage of this fact to extract information about the behavior
of an intermediate region of the gap when hopping is small but non-zero.

\begin{figure}[t]
\begin{center}
\includegraphics[width=1.0\linewidth]{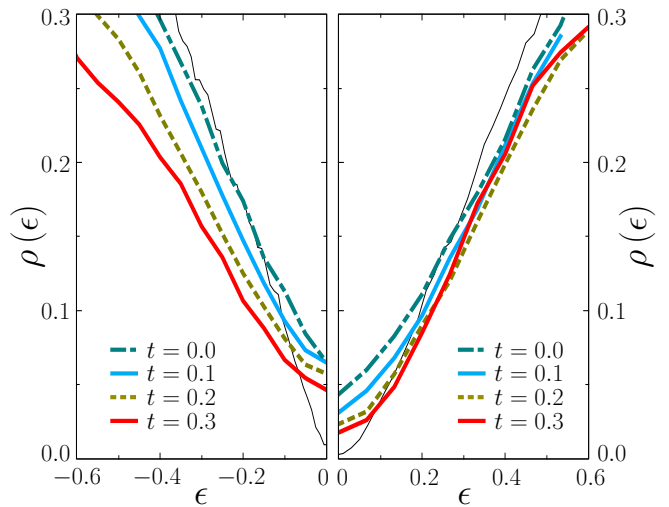}
\caption[]
 {(Color online). The DOS as a function of energy for size $L=20$ in 1D (left) and $N=4\times 4$ in 2D (right) systems for several values of hopping
as indicated in each panel.
The thin black lines represent much larger sizes for the classical case and are shown as references.
}\label{fig:Fig2}
\end{center}
\end{figure}

We study now the dependence of the DOS on the hopping parameter.
In the left panel of Fig.\ \ref{fig:Fig2} the DOS as a function of energy is shown for several 
values of the hopping parameter for a 1D system of size $L=20$.
From top to bottom, the different lines correspond to $t=0,\ 0.1,\ 0.2,\ 0.3$ respectively. The thin line is the classical DOS for a size $L=1024$ 
and it is included only as a reference.
We can see that the gap in the DOS corresponding to an intermediate region widens for finite values of the hopping parameter.
This is a surprising behavior according to the intuitive picture explained in the introduction, but it is similar to the 
results found in reference [\onlinecite{Lee99}] for 1D systems.
Another remarkable feature is that the shape of the gap remains pretty much linear 
for the values of $t$ shown. Near the Fermi level the gap is not linear, but one can ascribe this to finite size effects. 

The right panel of Fig.\ \ref{fig:Fig2} shows the DOS as a function of energy for $t=0,\ 0.1,\ 0.2,\ 0.3$ in the 2D case for size $N=4\times4$.
The thin black line corresponds to the classical DOS for a size $N=33\times 33$ 
and it is included as a reference.
Similarly to the 1D case, we can observe that the gap in the DOS widens at an intermediate energy region as $t$ increases up to $t=0.2$,
though the changes are smaller than in the 1D problem.
We stress that the tendency for $t=0.3$ seems to be the opposite that the one for smaller hopping.
That is, the DOS at $t=0.3$ seems to narrow respect the DOS at $t=0.2$.
These results are again compatible with reference [\onlinecite{Lee99}] but in contradiction
with all the other previous works treating the 2D case [\onlinecite{Vignale86},\onlinecite{Vignale87},\onlinecite{Li93}].

\section{Perturbation theory}\label{section:4}

Our direct diagonalization method was constrained to small systems.
In this section, we develop a novel approach based on perturbation of the hopping parameter which will allow us to calculate the DOS
for larger sizes and obtain reliable data for the DOS at Fermi level. 
Usually, perturbation theory is difficult to apply to the kind of interacting system we are treating.
In fact, a small perturbation of a hopping term between two given sites affects particles
which are arbitrarily far away of those sites due to the long-range character of the interaction.
Then, the perturbative analysis must be performed over the full many-body space instead of the one-particle states.
We have overcome these difficulties for small enough $t$ by only considering the most relevant hopping terms when computing the local DOS at a given site $i$.

In the classical case, $t=0$, all the eigenstates have a well defined occupation in each site.
As a consequence, there are well defined energies associated with each site of the lattice. To be more precise,
the energy of site $i$ is defined as:
\begin{align}\label{eq:site_energy}
\epsilon_{i}=
\begin{cases}
E^{p}_{i}-E_{0}, & \text{if }n_{i}=1\\
E_{0}-E^{h}_{i}, & \text{if }n_{i}=0
\end{cases},
\end{align}
where $n_{i}$ is the occupation of site $i$, $E_{0}$ the energy of the ground state and $E_{i}^{p},\ E^{h}_{i}$ are the energies of the 
eigenstates with a particle or hole respectively created at site $i$ to the ground state.
The DOS for a classical system is just the density of these site energies.
When quantum effects are introduced, this picture can be significantly altered. 
Indeed, the eigenstate of the Hamiltonian for non-zero hopping do not have a well defined occupation in each site and, 
in general, it is not possible to define a site energy. 
Nevertheless, we will see that a picture similar to the classical case still holds when the localization length is small enough.

We compute the DOS treating the hopping term as a perturbation to the classical Hamiltonian.
To do so, we can express the particle and hole DOS, Eqs. (\ref{eq:dp}) and (\ref{eq:dh}),  at site $i$ as a power series of the hopping $t$.
The first non-zero correction for these equations is of second order.
The particle DOS at site $i$ up to second order on the hopping parameter is:
\begin{align}\label{eq:Dp2}
 [\rho^{p}_{i}\left(t\right)](\epsilon) = &\  t^{2}\sum_{\alpha}     \delta \left( \epsilon - E^{p}_{\alpha}\left(0\right)+E_{0}\left(0\right) \right)  
                                              \frac{d^2}{dt^2}\bigg|_{t=0} A_{\alpha}^{i}(t) \nonumber\\
                                           & +\sum_{\alpha}   \delta \left( \epsilon - E^{p}_{\alpha}\left(t\right)+E_{0}\left(t\right) \right) A_{\alpha}^{i}(0) +
                                                  \mathcal{O}\left(t^{3}\right)
\end{align}
where $A_{\alpha}^{i}(t)=\left|\langle \psi_{\alpha}^{p}(t) |c_{i}^{\dag}|\psi_{0}(t)\rangle\right| ^{2}$ are 
the matrix elements appearing in Eq.\ (\ref{eq:dp})  as a function of the hopping.
The energy of the ground state up to second order correction on $t$ is $E_{0}\left(t\right)$
and $E_{\alpha}^{p}\left(t\right)$ is the energy of a state $\alpha$ with one more
particle than the ground state up to second order corrections on $t$.
An equivalent formula holds for the correction of the hole DOS Eq.\ (\ref{eq:dh}).
  
At second order on $t$, an electron can hop at most to a nearest neighbor site. 
Roughly speaking, an electron hopping from site $i+r$ to $i+r\pm 1$ changes the classical energy of the site $i$ 
as a dipolar contribution when $r$ is large enough.
Then, the changes in the classical energy at site $i$ due to the hopping between sites $i+r,\ i+r\pm1$ should fall off faster than
$1/|r|^{2}$. We compute the DOS including only the dominant hopping terms, that is, 
the hopping terms between the site where the particle is created (destroyed) and its nearest neighbors.
As we will see, this is an excellent approximation when hopping is small.

The first non-zero correction to the matrix elements $A_{\alpha}^{i}$ due to nearest neighbors hopping is:
\begin{align}\label{eq:Me}
A_{\alpha}^{i}(t)=A_{\alpha}^{i}(0)+t^{2}\sum_{j}C_{\alpha}^{i,j}+\mathcal{O}\left(t^{3}\right),
\end{align}
where the sum in $j$ runs over all nearest neighbors of site $i$ and the $C_{\alpha}^{i,j}$ are real constants. 
Using the standard perturbation theory, it is possible to prove that these constants fulfill $C_{\alpha}^{i,j}=-C_{\alpha}^{j,i}$.
Due to this property, the first term of 
the right hand side of Eq.\ (\ref{eq:Dp2})  cancels when the particle DOS is averaged over all the sites in a sample.
We can also simplify the second term. Indeed, the matrix element $A_{\alpha}^{i}(0)$ is non-zero only if the eigenstate $\alpha$ corresponds to
a state with an electron added at site $i$ on top of the ground state.
Then, our approximation leads to a DOS given by:
\begin{align}\label{eq:Ap_DOS}
\rho\left(\epsilon\right)=\langle\delta\left(\epsilon-\epsilon_{i}\left(t\right)\right)\rangle,
\end{align}
where $\epsilon_{i}(t)$ is the classical energy of site $i$, Eq.\ (\ref{eq:site_energy}), 
plus its second order correction due to nearest neighbors hopping
and the brackets mean the average over sites.
As in the classical case, each site contributes to the DOS in Eq.\ (\ref{eq:Ap_DOS}) with a well defined  site energy.

We have to determine the new site energies $\epsilon_{i}(t)$. 
In principle, the disorder breaks any degeneracy between the classical eigenstate and the standard-second order perturbation theory
could be employed to compute these energies.
However, the existence of classical eigenstates  with very similar energies can produce numerical instabilities.
To avoid this problem, for each pair of nearest neighbors sites $i$ and $j$ we use an exact diagonalization scheme equivalent to the perturbation of a degenerate level. 
This hopping term only produces corrections in the classical eigenstates when $n_{i}\neq n_{j}$.
Depending on the occupation of the ground state in these two sites, 
the hopping produces a correction either on the ground state or in a state with a particle or hole added on site $i$.
For each of these cases, the correction can be computed by one of the eigenvalues of a two level 
operator $H_{i,j}$.  
A compact way of expressing $H_{i,j}$ depending on the occupation and classical site energies of $i$ and $j$ is:
\begin{equation}\label{eq:Hamiltonian}
 H_{i,j}= \left(2n_{i}-1\right)\left( \begin{array}{ccc}
0     & -t \\
-t         & -d_{ij}
\end{array} \right)+\epsilon_{i}\ \mathbb{I},
\end{equation}
where the factor $\left(2n_{i}-1\right)$ is introduced to match the sign of the energy correction for particles (positive) and holes (negative) 
and 
\begin{align}\label{eq:dfactor}
d_{ij}=
\begin{cases}
\epsilon_{i}- \epsilon_{j}, & \text{if }n_{i}=n_{j}\\
-\epsilon_{i}+ \epsilon_{j}+1& \text{if }n_{i}<n_{j}\\
-\epsilon_{j}+ \epsilon_{i}+1& \text{if }n_{i}>n_{j}
\end{cases}.
\end{align}
Finally, the site energy of $i$ is the eigenvalue of $H_{ij}$ which reduces to $\epsilon_{i}$ when $t=0$. 

\begin{figure}[t!]
\begin{center}
\includegraphics[width=1.0\linewidth]{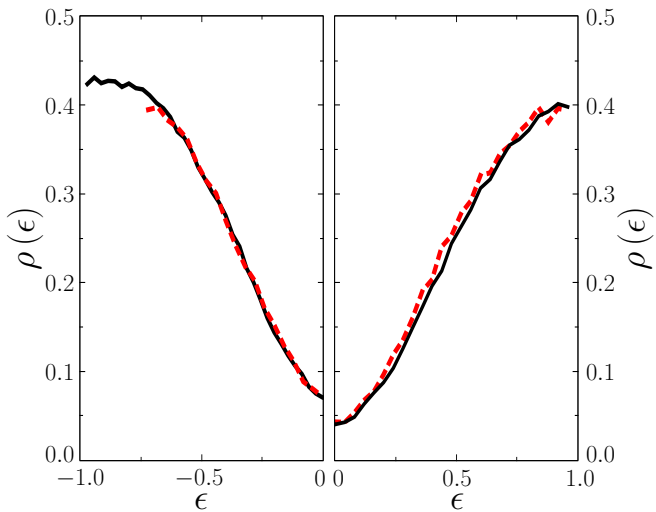}
\caption[]
{The DOS as a function of the energy for $t=0.1$ computed with the perturbation method (black solid line) 
and with exact diagonalization (red dashed line) for $L=16$ in 1D systems (left) and $N=4\times 4$ in 2D  systems (right).
}\label{fig:Fig3}
 \end{center}
\end{figure}

We sum all the corrections given by $H_{i,j}$ for $j$ being a nearest neighbor of site $i$. 
This procedure can be viewed as a partial re-summation of the full perturbation serie in order to avoid 
numerical instabilities, although it is only exact up to the lowest non-zero order.
Then, we can express the site energy including all these corrections as:
\begin{align}\label{eq:site_energy_Dia}
\epsilon_{i} (t)=\epsilon_{i}+\left( n_{i}-\frac{1}{2}\right)\sum_{j}d_{ij}\left(1-\sqrt{1+\left(\frac{2\ t}{d_{ij}}\right)^{2}}\right),
\end{align}
where the summation in $j$ runs again over nearest neighbors of site $i$

We notice that the ground state for non-zero hopping $t$ can be different from the perturbed classical ground state.
Indeed, some other low-energy state is likely to have the smallest energy when a finite hopping is taken into account.
To deal with this problem we compute the quantum corrections of the first $S$ classical states
and we chose as the new ground state the state with the minimum total energy.
We found that $S=10000$ is enough 
for system sizes up to $L\sim 300$ in 1D and up to $N\sim 10\times  10$ in 2D systems for the values of $t$ considered. 
Then, the site energies given by Eq.\ (\ref{eq:site_energy_Dia}) are computed using the $d_{i,j}$ factors associated to the zero order of this
new ground state.

\subsection{Results}

\begin{figure}[t!]
\begin{center}
\includegraphics[width=1.0\linewidth]{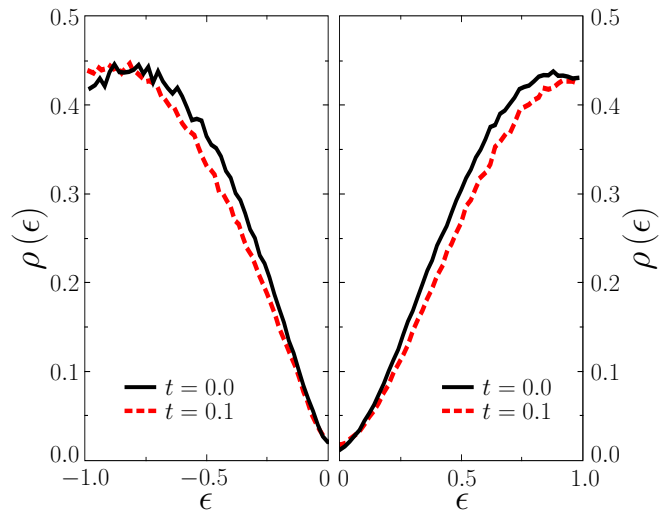}
\caption[]
{The DOS as a function of the energy for $t=0$ (black solid line)
and $t=0.1$ (red dashed line) for 1D systems of size $L=256$ (left) and 2D systems of size $N=10\times 10$ (right) 
both of them computed with the perturbation method.
}\label{fig:Fig4}
 \end{center}
\end{figure}

We begin checking the validity of our approximation by comparing with
the exact DOS.
In the left panel of Fig.\ \ref{fig:Fig3}, the DOS computed with exact diagonalization and the result from perturbation theory
are shown as a function of the energy  for 1D systems for hopping 
$t=0.1$ and  size $L=16$. 
At the right panel, the perturbed and exact DOS for 2D systems are presented for hopping $t=0.1$ and size $N=4\times 4$. 
As we still be at half filling where the DOS is symmetric respect the Fermi energies,
only half of the DOS is shown as we did in the previous section.
The agreement between both methods is remarkably good in 1D and 2D systems.
For 1D systems, we have checked that the perturbation method gives sensible results even for $t=0.2$,
while for 2D systems the results are not so good.
We have also measured the localization length in 1D systems obtaining $\xi(t=0.1)\approx 0.8$
and $\xi(t=0.2)\approx 1.4$ [\onlinecite{Israel}]. 
We conclude that the perturbation method employed is quite accurate for small values of $\xi$, of the order of the lattice spacing.
In higher dimensions $\xi$ increases faster than in 1D systems, thus, the range of validity of the approach is smaller.

The DOS  as a function of the energy for $t=0$ and $t=0.1$ is shown using 1D systems of sizes $L=256$ (left panel of Fig.\ \ref{fig:Fig4}) and in 
2D systems of sizes $N=10\times 10$ (right panel of Fig.\ \ref{fig:Fig4}).
First, as the sizes are much larger than the one used for the exact diagonalization method, the finite size effects are also much smaller.
So now, we can extract a more reliable information about the behavior of the DOS near the Fermi energy.
In Fig.\ \ref{fig:Fig4}, we can appreciate  that the DOS near this Fermi level does not seem to be affected by a small hopping in 1D neither in 2D systems.
Indeed, we have represented the DOS at zero energy, $\rho\left(\epsilon=0\right)$, as a function of the inverse of the linear size, $1/L$, for 2D system of linear sizes
$L=4,6,8,10$ finding that the results are compatible with a $\rho\left(\epsilon=0\right)=0$ at the thermodynamical limit.
On the other hand, we can appreciate in Fig.\ \ref{fig:Fig4} that the DOS widens for a intermediate energy region as it was also found using  exact diagonalization. 
Thus, the results indicate that the inclusion of  a weak quantum perturbation reduces the slope of 
the Coulomb gap in 1D and 2D systems. 

\begin{figure}[t!]
\begin{center}
\includegraphics[width=1.0\linewidth]{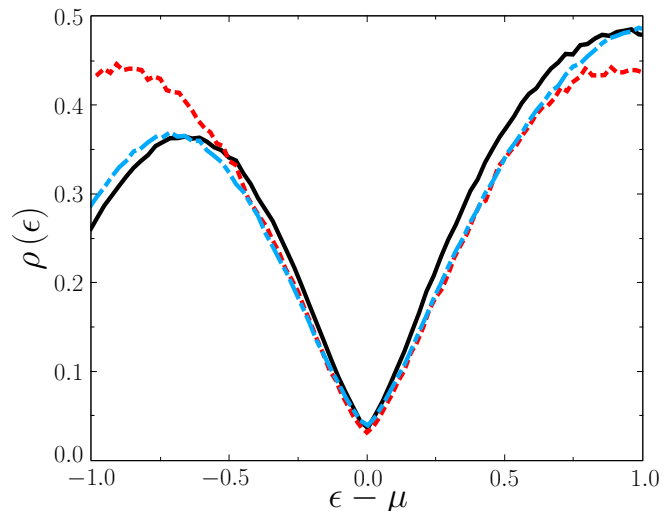}
\caption[]
{(Color online).The red dashed line represents the DOS as a function of the energy for $t=0.1$ at half filling ($\mu=0$) for 1D systems of size $L=90$
computed with the perturbation method.
The blue semi-dashed line is the DOS for $t=0.1$ obtained also with the perturbation method and the black solid line is the classical DOS, $t=0$, 
 both of them computed for size $L=90$ at one third filling where the compensation constant is $K=1/3$
and the chemical potential $\mu=-1/3$.  
}\label{fig:Fig5}
 \end{center}
\end{figure}

Finally, we check that our findings do not depend on the degree of filling. We compute the DOS at one third filling 
where the compensation constant is chosen as $K=1/3$ and chemical potential $\mu=-1/3$.
In Fig.\ \ref{fig:Fig5}, we can see the DOS for hopping values $t=0$ (black solid line) and $t=0.1$ (blue semi-dashed line) as a function of the energy calculated
for a 1D system of size $L=90$ at one third filling and setting the Fermi level at the origin of energies.
We also compare the DOS at this filling with the one that has already been calculated for half filling.
The red dashed line is the DOS for $t=0.1$ as a function of the energy at half filling for a system of size $L=90$.
We see that the DOS for hopping $t=0.1$ at both fillings have the same shape near each of its respective Fermi energies. We have checked that the same
result holds in 2D.
Then, the shape of the DOS near the Fermi energy for a small but non-zero hopping is independent of 
the compensation as it also occurs with the classical Coulomb gap [\onlinecite{Efros76}].

\section{Conclusions}\label{section:5}

Our numerical results indicate that weak quantum effects produce a widening of the Coulomb gap in the DOS.
This widening is due to level repulsion.
In the ground state, when two nearest-neighbor sites have opposite occupancy, the hopping term mixes two configurations: 
$\left[ \circ \bullet\right]$ and $\left[ \bullet \circ\right]$. Due to level repulsion the difference 
in energy must be larger than $t$. The energy of the ground state is always reduced  by  quantum corrections, while  configurations with one more or one less 
particle are not affected (up to $\mathcal{O}(t^2)$ ). 
This effect contributes to open the gap, as occupied sites will reduce its energy while 
empty ones will increase it. If two nearest sites are both empty in the ground state $\left[ \circ \circ\right]$,
quantum corrections will affect only configurations with one more particle. 
Indeed, the site with lowest site energy will reduce it (due to level repulsion) and the other one will increase it. 
So, for an empty site near the center of the gap this second effect will produce, on average, a reduction of the site energy (the gap closes) 
as the chances are that the empty neighbor will have larger site energy. For 1D and 2D we expect this second effect to be much smaller than 
the first one because the probability of finding two empty sites with site energies within the gap and difference smaller than $t$ is very small. 
The first effect does not require that both site energies should be inside the gap, so it should be more likely.
In 3D systems, as the gap is parabolic, this second mechanism is more effective, so we expect the widening of the gap to be much smaller.

The previous arguments are not rigorous as quantum effects, for a finite $t$, could change the 
configuration of the ground state. More properly we should use a self-consistent approximation
taking into account the quantum corrections. 
The net effect of this quantum corrections is the hardening of the constraints for the ground state. In particular  an occupied site with energy $\epsilon_i(t)$
and a nearest neighbor empty site with energy $\epsilon_j(t)$ must satisfy the condition
\begin{equation}\label{eq.splitting}
\epsilon_{j}(t)-\epsilon_{i}(t)-1\ge t.
\end{equation}

We may notice that the widening of the Coulomb gap with transfer energy $t$, specially on low dimensional systems,
should be taken into account in the interpretation of tunneling experiments in the strongly localized regime [\onlinecite{Massey_Lee_Shklovskyi},\onlinecite{experimental_gap2}].
It may also affect variable range hopping conductance. 
A calculation based on the classical DOS [\onlinecite{TsEf02},\onlinecite{conduct}] will overestimate the conductance, since the widening of the gap will not be taken into account.

In summary, we have studied quantum effects in the DOS of strongly localized regime with a $1/\sqrt{r}$ interaction in one dimension and a Coulomb interaction in two dimensions.
Two different numerical methods have been employed.
First, an exact diagonalization algorithm has allowed us to compute the DOS for small sizes.
We have obtained that the DOS in one and two dimensions remains approximately linear for small but non-zero hopping.
For 1D systems, a monotonous reduction in the slope of the DOS near the Fermi level has been found when hopping increases from $t=0$ to $t=0.3$. 
For 2D systems, the Coulomb gap also widens when the hopping is increased from $t=0$ to $t=0.2$, 
but this tendency is reversed from there on. 
Using a novel perturbation  approach, we have been able to compute the DOS for much larger system sizes than for exact diagonalization
using hopping $t=0.1$. The results confirm that the slope of the Coulomb gap decreases when including weak quantum effects 
and that there is not an appreciable filling of the Coulomb gap at the Fermi energy.
Finally,  we have computed the DOS at $1/3$ filling for 1D systems. Our result indicates that the behavior of the DOS near the Fermi energy 
is independent of the degree of filling.

\acknowledgements
This work was supported by the DGI Grant No. FIS2009-13483 of the Spanish Ministerio de Econom\'{i}a y Competitividad.
We thank the Centro de Supercomputaci\'on de la Fundaci\'on Parque Cient\'{i}fico de Murcia where part of the numerical work was performed.

\end{document}